\documentstyle[12pt]{article}

\def\Z {{\bf Z}}
\def\R {{\bf R}}
\def\C {{\bf C}}
\def\H {{\bf H}}
\def\J {{\bf J}}
\def\B {{\bf B}}
\def\K {{\bf K}}

\newcommand{\wh}{\widehat}
\newcommand{\ol}{\overline}
\newcommand{\ra}{\rightarrow}

\newcommand{\cA}{{\cal A}}
\newcommand{\cB}{{\cal B}}
\newcommand{\cH}{{\cal H}}
\newcommand{\cF}{{\cal F}}
\newcommand{\del}{\partial}
\newcommand{\sq}{\sqrt{2}}

\def\diag{\mathop{\rm diag}\nolimits}

\def\mod{\mathop{\rm mod}\nolimits}

\makeatletter
\@addtoreset{equation}{section}
\makeatother

\def\mat#1{\matt[#1]}
\def\matt[#1,#2,#3,#4]{\left(%
\begin{array}{cc} #1 & #2 \\ #3 & #4 \end{array} \right)}

\def\v2#1{\vv2[#1]}
\def\vv2[#1,#2]{\left(%
{#1 \atop #2}\right)}

\def\drawbox#1#2{\hrule height#2pt
        \hbox{\vrule width#2pt height#1pt \kern#1pt
              \vrule width#2pt}
              \hrule height#2pt}

\def\Fund#1#2{\vcenter{\vbox{\drawbox{#1}{#2}}}}
\def\Asymm#1#2{\vcenter{\vbox{\drawbox{#1}{#2}
              \kern-#2pt       
              \drawbox{#1}{#2}}}}

\def\fnd{\Fund{5.5}{0.4}}
\def\asym{\Asymm{5.5}{0.4}}
\def\sym{\fnd\kern-0.4pt\fnd}

\begin{document}

\begin{titlepage}
\vspace*{-2.5cm}
\null
\begin{flushright} 
hep-th/0202165 \\
YITP-02-11\\
CITUSC/02-005\\
ITFA-2002-04\\
February, 2002
\end{flushright}
\vspace{0.5cm} 
\begin{center}
{\Large \bf
D-branes and KK-theory in Type I String Theory
\par}
\lineskip .75em
\vskip1.5cm
\normalsize

{\large
 Tsuguhiko Asakawa\footnote{
E-mail:\ \ asakawa@yukawa.kyoto-u.ac.jp },
 Shigeki Sugimoto\footnote{
E-mail:\ \ sugimoto@citusc.usc.edu}  and 
Seiji Terashima\footnote{
E-mail:\ \ sterashi@science.uva.nl} }
\vskip 2.5em
{
${}^1$
\it
Yukawa Institute for Theoretical Physics,
Kyoto University,\\
Kyoto 606-8502, Japan
}
\vskip 1em
{
${}^2$
 \it
CIT/USC Center for Theoretical Physics,
University of Southern California,\\
 Los Angeles CA90089-2535, USA \\}
\vskip 1em
{
${}^3$
 \it  Institute for Theoretical Physics,
University of Amsterdam \\
Valckenierstraat 65, 
1018 XE, Amsterdam, The Netherlands}
\vskip 2em

{\bf Abstract}

\end{center}
We analyse unstable D-brane systems in type I string theory.
Generalizing the proposal in hep-th/0108085, we
give a physical interpretation for real KK-theory
and claim that the D-branes embedded in
a product space $X\times Y$ which are
made from the unstable D$p$-brane system wrapped on $Y$
are classified by a real KK-theory group $KKO^{p-1}(X,Y)$.
The field contents of the unstable D-brane systems
are systematically described by a hidden Clifford algebra
structure.

We also investigate the matrix theory based on
non-BPS D-instantons and show that the spectrum of
D-branes in the theory is exactly what we expect in
type I string theory, including stable non-BPS D-branes
with $\Z_2$ charge.
We explicitly construct the D-brane solutions
in the framework of BSFT and analyse the physical
property making use of the Clifford algebra.

\end{titlepage}

\baselineskip=0.7cm

\section{Introduction}

It has been argued that
the D-brane charges are classified by K-theory
\cite{MiMo,Wi,Ho}.
Each element of K-theory is interpreted as
a configuration of the gauge bundle and tachyon fields on
a space-time filling unstable D-brane system,
such as non-BPS D9-branes in type IIA \cite{Ho}
or D9-brane - anti D9-brane
system in type IIB \cite{Wi}, describing a set of
lower dimensional D-branes. This interpretation is
obtained by generalizing
the D-brane descent relations discussed in
\cite{Se}.

On the other hand, it is also known that
we can construct higher dimensional D-branes from
a lower dimensional D-brane system \cite{To,BFSS,IKKT,Is,Kl,Te}.
Therefore, it is natural to ask how we can classify the
D-branes made by a lower dimensional unstable D-brane system.
In the previous paper \cite{AsSuTe}, we argued that
K-homology groups are the groups which classify D-branes
in matrix theories based on non-BPS D-instantons in type IIA
or D-instanton - anti D-instanton system in type IIB,
which we called K-matrix theory.
The K-homology is a dual of K-theory and it turns out that this
result is consistent with the classification of D-brane charge
using K-theory.
This means that the K-matrix theory correctly reproduces the 
D-brane spectrum expected from K-theory.

Furthermore, the argument is
generalized to higher dimensional
systems and we found that Kasparov's KK-theory groups 
are the relevant groups
for the classification.
To be more precise, suppose that
the space-time manifold is a product space $X\times Y$,
and consider stable D-branes made by the unstable D-brane
system wrapped along $Y$. Then, the D-branes are naturally classified
by the KK-theory group denoted by $KK^i(X,Y)$.
(See section \ref{KKcomplex} for a brief review.)
This group generalizes the above results.
Actually, when the space $X$ or $Y$ is a point, the KK-theory group
reduces to K-theory or K-homology, respectively.
It is also shown in \cite{AsSuTe} that the spectrum of the D-branes
does not depend on the choice of the unstable D-brane system,
using some isomorphisms among the KK-theory groups.

In this paper, we generalize the argument to type I string theory
and consider type I K-matrix theory, i.e. the matrix theory based on
an infinite number of non-BPS D-instantons in type I string theory,
as well as the other unstable D-brane systems in type I string theory.
We are particularly interested in the construction of D-branes
in these systems.
As shown in \cite{Se,Wi,Gu}, type I string theory has
non-trivial D-brane spectrum even when the
space-time manifold is flat. In fact, the charge of
flat D$p$-branes transverse to $\R^{9-p}$ is classified by
the real K-theory group $KO(\R^{9-p})$
\footnote{Here $KO(\R^{n})$ denotes
the reduced K-theory group of $S^n$, $\widetilde{KO}(S^n)$
\cite{Kar}.}
\cite{Wi}, which is given in table \ref{spec}.
\begin{table}[htb]
\begin{center}
$$
\begin{array}{c|ccccccccccc}
\hline\hline
p&-1&0&1&2&3&4&5&6&7&8&9\\
\hline
KO(\R^{9-p})&\Z_2&\Z_2&\Z&0&0&0&\Z&0&\Z_2&\Z_2&\Z\\
\hline
\end{array}
$$
\parbox{75ex}{
\caption{ The spectrum of flat D$p$-branes in type I theory,
which is classified by the real K-theory group $KO(\R^{9-p})$.
\small
}
\label{spec}
}
\end{center}
\end{table}
In particular, there are stable non-BPS D-branes with $\Z_2$ charges.
So, it would be more challenging to explore this theory than
the type II sting theory.
We will show that we correctly obtain this spectrum using
type I K-matrix theory.

It is easy to generalize the idea to
higher dimensional systems. Actually, we will not
restrict our arguments to the matrix theory, but consider
D-branes made by unstable D$p$-brane
systems in type I string theory, (i.e.
 non-BPS D$p$-branes for $p=-1,0,2,3,4,6,7,8$
and D$p$-brane - anti D$p$-brane system for $p=1,5,9$).
The gauge groups and the representation of tachyon fields
on the world-volume of non-BPS D$p$-branes are
examined in \cite{Be}, and the results
are summarized in table \ref{unstable}.
\footnote{
In this table, we omitted
the massless scalar and tachyon modes from
D$p$-D9 strings for $p\ge 5$, since
they do not affect the K-theory classification,
as discussed in \cite{Be,LoUr}.
So, we simply neglect them in this paper.}
It is also shown in \cite{Be,Wi}
that the K-theory group which
classifies the charge of D-branes made by the descent relations
from the unstable D$p$-brane system
is the real K-theory group $KO^{p-1}(Y)$
($\simeq KO^{p-9}(Y)$),
where $Y$ is the world-volume manifold of the system.
\begin{table}[htb]
\begin{center}
$$
\begin{array}{c|cccl}
\hline\hline
p&\mbox{Gauge}&\mbox{Tachyon}&\mbox{Scalar}&\mbox{K-group}\\
\hline
-1&U&\asym&\mbox{adj.}&KO^{-2}(pt)\\
0&O&\asym&\sym&KO^{-1}(Y)\\
1&O\times O&(\fnd,\fnd)&(1,\sym),(\sym,1)&KO^{0}(Y)\\
2&O&\sym&\sym&KO^1(Y)\\
3&U&\sym&\mbox{adj.}&KO^2(Y)\\
4&Sp&\sym&\asym&KO^3(Y)\\
5&Sp\times Sp&(\fnd,\fnd)&(1,\asym),(\asym,1)&KO^4(Y)\\
6&Sp&\asym&\asym&KO^{-3}(Y)\\
7&U&\asym&\mbox{adj.}&KO^{-2}(Y)\\
8&O&\asym&\sym&KO^{-1}(Y)\\
9&O\times O&(\fnd,\fnd)&--&KO^0(Y)\\
\hline
\end{array}
$$
\parbox{75ex}{
\caption{
\small The gauge groups, tachyons and scalar fields on
the world-volume of type I unstable
D$p$-brane systems and corresponding K-theory groups.
}
\label{unstable}
}
\end{center}
\end{table}

{}From the argument analogous to that given in type II case,
it is quite natural to expect that
the classification of type I D-branes via the real K-theory
is generalized by using real KK-theory,
when we take into account
the D-branes stretched along the directions transverse to the unstable
D$p$-brane system.
As we will explain in section \ref{KKO},
the relevant KK-theory group
which generalizes the K-theory group $KO^{p-1}(Y)$ is
the real KK-theory group $KKO^{p-1}(X,Y)$, where $X$ is
the manifold transverse to the world-volume of the unstable D$p$-brane
system wrapped along $Y$.
Note that when we forget about the transverse space $X$
by setting $X$ to be a point,
the KK-theory group $KKO^{p-1}(pt,Y)$
reduces to the real K-theory group $KO^{p-1}(Y)$
and we correctly recover the K-theory results, though
we describe the K-theory group using Fredholm operators
as in \cite{HaMo,Wi2}, which is slightly different
from the description used in \cite{Be}.
We will give a physical interpretation for each element of the
KK-theory group in terms of the world-volume theory
on the unstable D$p$-brane system.
As one can see from the table,
the world-volume theory changes drastically as the dimension
of the system changes.
Therefore, it provides quite non-trivial check for our
interpretation of the elements of $KKO$ groups in terms
of the world-volume field theory.

The paper is organized as follows.
In section \ref{KKandD}, we first review complex KK-theory and
its physical interpretation in terms of unstable D-brane
systems in type II theory \cite{AsSuTe},
and explain our proposal for type I theory using the real KK-theory.
The physical interpretation of the elements of real KK-theory
is given in section \ref{realKKvsD}. We argue that
the field contents of the unstable D-brane system listed
in table \ref{unstable} nicely fit in with the
definition of the real KK-theory groups.
As we will see, the contents of table \ref{unstable}
can be easily reproduced by looking at real Clifford algebras,
which are used in the definition of the KK-theory groups.
Section \ref{flat} is mainly devoted to analyse
type I K-matrix theory.
We will explicitly construct flat D$p$-brane solutions
in the framework of BSFT \cite{KuMaMo,KrLa,TaTeUe}
for the type I non-BPS D-instanton system.
We examine the tension of the D-brane solutions
and tachyon modes on them, making use of the Clifford algebra
structure, and reproduce the expected property.
Finally, we discuss further applications, such as
the description of Chern-Simons terms using real superconnections,
in section \ref{Disc}.

\section{KK-theory and D-branes}
\label{KKandD}
\subsection{Complex KK-theory and type II D-branes}
\label{KKcomplex}

First, we sketch the physical interpretation
of complex KK-theory given in \cite{AsSuTe}.

Let us consider type II string theory on a product space
$X\times Y$. As argued in \cite{Se,Wi,Ho},
D-branes can be constructed as solitons
on space-time filling unstable D-brane systems, such as 
non-BPS D9-branes in type IIA or D9-brane - anti D9-brane
system in type IIB, wrapped on $X\times Y$.
On the other hand, as mentioned in the introduction,
we can construct higher dimensional
D-branes from a lower dimensional unstable D$p$-brane system.
In particular, we can construct the space-time filling unstable
D-brane system, as well as the D-brane solitons on it, from
the lower dimensional unstable D$p$-brane system.
Therefore, any D-brane configurations can in principle
be represented by the unstable D$p$-brane system.
So, let us suppose the dimension of $Y$ is $p+1$ and
consider unstable D$p$-brane system wrapped on $Y$.
Then, stable D$q$-branes ($q\le p$) inside $Y$ are contained
as solitons, which represents a K-theory class of $Y$,
in the system. Moreover, D-branes wrapped on a subspace
of the transverse space $X$ can also be constructed.
For example, if we are interested in the configurations
which are constant along $Y$, the construction of such
D-branes is the same as that given
in the K-matrix theory \cite{AsSuTe,Te}.
Note that this construction is similar to those given
in supersymmetric matrix theories \cite{To,BFSS,IKKT}, but here
the tachyon fields play an essential role.
These configurations
are naturally classified by the analytic K-homology of $X$
\cite{AsSuTe}.
Therefore,
in order to classify the D-brane configurations
in the unstable D$p$-brane system,
we need a mathematical framework which generalizes
both K-theory and K-homology.

The KK-theory is a generalization of both K-theory and K-homology,
introduced by Kasparov \cite{Kas,Bl}.
There are two kinds of complex KK-theory groups denoted by
$KK^i(X,Y)=KK_i(C_0(X),C_0(Y))$ ($i=0,1$). Here $X$ and $Y$ are
topological spaces (locally compact Hausdorff spaces), and
$C_0(X)$ denotes the set of complex valued continuous functions on $X$
vanishing at infinity.
\footnote{
Note that every commutative $C^*$-algebra can be
expressed as $C_0(X)$. One can define
$KK_n(\cA,\cB)$ for arbitrary (could be non-commutative)
$C^*$-algebras $\cA$ and $\cB$. The generalization to
non-commutative cases is straightforward, though
the physical interpretation is unclear in general.
}

The KK-theory group $KK^1(X,Y)$
\footnote{Here we assume that $X$ and $Y$ are compact, for simplicity.}
is defined as the set of equivalence classes of triples $(\cH,\phi,T)$,
called odd Kasparov modules, where
$\cH$ ($= C_0(Y)^\infty$) is a Hilbert space
over $C_0(Y)$,
\footnote{
Hilbert space over a $C^*$-algebra $\cA$, denoted
by $\cA^\infty$, is a Hilbert $\cA$-module defined as
$\cA^\infty=\{
(x_k)\in\prod_{n=1}^\infty\cA\,|\, \sum_k x_k^* x_k~~ \mbox{converges
 in $\cA$}\}$.
}
$\phi: C_0(X)\ra \B(\cH)$  is a *-homomorphism
and $T$ is a self-adjoint operator in $\B(\cH)$ such that
\begin{eqnarray}
T^2-1 \in \K(\cH),~~
{[}T,\phi(a){]} \in \K(\cH)~~\mbox{for}~{}^\forall a\in C_0(X).
\label{KK1}
\end{eqnarray}
Here $\B(\cH)$ is the set of adjointable operators on $\cH$
\footnote{An operator $T:\cH\ra\cH$ is called adjointable if
there is an operator $T^*:\cH\ra\cH$ with
$\langle Ta,b \rangle=\langle a,T^*b \rangle$
for all $a,b\in\cH$. Here $\langle a,b \rangle=\sum_n a_n^*b_n$
for $a=(a_n),b=(b_n)\in\cH$.
Adjointable operators on a Hilbert $\cA$-module are automatically
$\cA$ module homomorphism.
}
and $\K(\cH)$ is the closure of `finite rank' operators in $\B(\cH)$.
\footnote{
A `finite rank' operator is
a linear span of the operators $\theta_{x,y}$ defined
as $\theta_{x,y} z= x\langle y,z \rangle$ for $x,y,z\in\cH$
}
Note that when $Y$ is a point, $\cH$ is a separable Hilbert space
over $\C$, $\B(\cH)$ is the set of bounded linear operators on the
Hilbert space $\cH$, and $\K(\cH)$ is the set of compact operators on
$\cH$.

The equivalence relations are defined as follows.
Two Kasparov modules ($\cH_i,\phi_i,T_i$) ($i=0,1$)
are called unitary equivalent when there is a unitary operator $U$ in
$\B(\cH_0,\cH_1)$ such that $T_0=U^*T_1U$ and
$\phi_0(a)=U^*\phi_1(a)U$ for all $a\in C_0(X)$.
They are called operator homotopic if $\cH_0=\cH_1$, $\phi_0=\phi_1$ and
there is a norm continuous path between $T_0$ and $T_1$.
We define a degenerate Kasparov module
as the Kasparov module $(\cH',\phi',T')$
satisfying ${T'}^2-1= [T',\phi'(a)] =0$ 
for all $a\in C_0(X)$.
Then, in the definition of the KK-theory group,
 the two Kasparov modules
 ($\cH_i,\phi_i,T_i$) ($i=0,1$) are defined to be equivalent
if there are degenerate Kasparov modules
$(\cH'_i,\phi'_i,T'_i)$ ($i=0,1$) such that
$(\cH_i\oplus \cH'_i,\phi_i\oplus\phi'_i,T_i\oplus T'_i)$
($i=0,1$) are operator homotopic up to unitary equivalence.

This group classifies the solitonic configurations,
which turn out to be D-branes embedded in the space-time
$X\times Y$, in the system of an infinite number of non-BPS D$p$-branes
extended along $Y$ and perpendicular to $X$.
The space $\cH$ is identified as
a set of global sections of the infinite rank Chan-Paton bundle
associated with the non-BPS D$p$-branes.
The unitary transformation
acting on $\cH$ is nothing but the gauge transformation of the system.
The operator $T$ is interpreted as the tachyon field on
the non-BPS D$p$-branes.
The scalar fields $\Phi^i$
on the non-BPS D$p$-branes, which represents their transverse
positions, correspond to the operator
$\phi(x^i)$, which are the image of the coordinate functions $x^i$
under the *-homomorphism $\phi$.
Note that
there are some delicate issues for the choice of
the coordinate functions $x^i$,
which we won't explain in detail here. (See \cite{AsSuTe}.)
All we need in the following is the fact that
the scalar fields $\Phi^i$ are self-adjoint operators
in the image of the *-homomorphism $\phi$.

The condition (\ref{KK1}) is related to the finiteness
of the action. Here the tachyon is normalized such that
the minimum of the potential is given by $T^2=1$. Hence, roughly
speaking, the condition (\ref{KK1}) represents that almost all
the non-BPS D$p$-branes are annihilated.
 (See \cite{AsSuTe} for more detail.)

The equivalence relations also have a nice physical interpretation.
The unitary equivalence is nothing but the
gauge equivalence, and the operator homotopy
is just a continuous deformation of the tachyon configuration.
The degenerate elements are
interpreted as non-BPS D-instantons that would be
annihilated by the tachyon condensation.

When $X$ is a point,
the KK-theory group $KK^1(pt,Y)$
is isomorphic to the K-theory group $K^1(Y)$.
Therefore, we correctly reproduce the K-theory
classification,
if we are not interested in the the space $X$ which is
transverse to the non-BPS D$p$-branes.
In fact, the above interpretation
reduces to that given by Witten in \cite{Wi2}.
Similarly, when $Y$ is a point,
the KK-theory group $KK^1(X,pt)$
is isomorphic to the K-homology group $K_1(X)$,
which is consistent with the classification of D-brane
configurations in the K-matrix theory \cite{AsSuTe}.

The other KK-theory group $KK^0(X,Y)=KK(X,Y)$ consists of
the equivalence classes of
$\Z_2$ graded triples
$(\wh\cH,\wh\phi,F)$, called even Kasparov modules,
where
\begin{eqnarray}
\wh\cH=\v2{\cH^{(0)},\cH^{(1)}},
~~\wh\phi=\mat{\phi^{(0)},,,\phi^{(1)}},
~~F=\mat{,T^\dag,T,}.
\label{ev}
\end{eqnarray}
Here $\cH^{(i)}$ ($i=0,1$) are Hilbert spaces over $C_0(Y)$,
$\phi^{(i)}: C_0(X)\ra \B(\cH^{(i)})$ ($i=0,1$) are *-homomorphisms,
and $T:\cH^{(0)}\ra\cH^{(1)}$ is an adjointable operator satisfying
\begin{eqnarray}
F^2-1 \in \K(\wh\cH),~~
{[}F,\wh\phi(a){]} \in \K(\wh\cH)~~\mbox{for}~{}^\forall a\in C_0(X).
\label{KK0}
\end{eqnarray}
The grading is defined by the operator $\gamma=\diag(1,-1)$.
Note that the operator $\wh\phi(a)$ and $F$ are chosen to
be even and odd under this grading, respectively.
The equivalence relation for $KK^0(X,Y)$ is defined
in an analogous way as above. Note that the unitary operators,
which are used in the definition of unitary equivalence,
are required to be even with respect to the $\Z_2$ grading.

The physical interpretation of $KK^0(X,Y)$
is quite analogous to that given for
$KK^1(X,Y)$. Here $KK^0(X,Y)$ classifies D-branes made by
infinitely many D$p$-brane - anti D$p$-brane pairs wrapped on $Y$.
$\cH^{(0)}$ and $\cH^{(1)}$ are interpreted as the infinite dimensional
Chan-Paton Hilbert spaces of D$p$-branes and anti D$p$-branes, respectively,
and $T$ is the tachyon field created by the D$p$ - anti D$p$
strings.

It is worth noting that the odd Kasparov module
$(\cH,\phi,T)$ is also written as the even Kasparov module by
setting $\cH^{(0)}=\cH^{(1)}=\cH$,
$\wh\phi=\phi\otimes 1_2$ and $F=T\otimes \sigma_1$.
In fact, these two KK-theory groups are related by the isomorphism
\footnote{ Here we assume that the $C^*$-algebras $\cA$ and $\cB$
are trivially graded. For general $\Z_2$-graded $C^*$-algebras,
we should replace the tensor product $\otimes$ with
the graded tensor product denoted by $\wh\otimes$. (See \cite{Bl,Sc}.)}
\begin{eqnarray}
KK_1(\cA,\cB)\simeq KK_0(\cA,\cB\otimes \C_1),
\label{equivKK}
\end{eqnarray}
where
$\C_1=\C\oplus \C\,e_1$
is a (complex) Clifford algebra generated by an element $e_1$
satisfying $e_1^2=1$, which can be represented by
associating $e_1$ to one of the Pauli matrices $\sigma_1$.
A Clifford algebra has a natural ($\Z_2$-) grading which is
given by regarding each generator as an odd element.
The elements of
$KK_0(C_0(X),C_0(Y)\otimes \C_1)$ is written as
$(\cH\otimes 1\oplus\cH\otimes e_1,\phi\otimes 1,T\otimes e_1)$,
where $(\cH,\phi,T)$ is an odd Kasparov module.


\subsection{Real KK-theory and type I D-branes}
\label{KKO}

The real KK-theory is defined in the same way as above, except
that we should use real objects, such as
real $C^*$-algebras, real Hilbert spaces,
real Clifford algebra, and so on. (See for example \cite{Sc}.)
The analog of $KK(X,Y)$ ($=KK^0(X,Y)$) is denoted by
 $KKO(X,Y)=KKO(C_0(X,\R),C_0(Y,\R))$, where $C_0(X,\R)$ is
a set of continuous real functions on $X$ vanishing at infinity.
The KK-theory groups $KKO^{-n}(X,Y)$, which we mainly consider,
are defined by using the Clifford algebra.
The Clifford algebra $\C^{p,q}$ is defined as an algebra over $\R$
generated by $e_i$ ($i=1,\dots,p+q$) satisfying
\begin{eqnarray}
&e_ie_j+e_je_i=0& (i\ne j)\\
&e_i^2=-1& (i=1,\dots,p)\\
&e_i^2=1& (i=p+1,\dots,p+q).
\end{eqnarray}
(See Appendix \ref{App1} for some more details about the Clifford algebra.)
And we define
\begin{eqnarray}
KKO_{q-p+r-s}(\cA,\cB)=
KKO(\cA\otimes \C^{p,q},\cB\otimes \C^{r,s}).
\label{KKO_n}
\end{eqnarray}
Note that one can show that the right hand side depends
only on $q-p+r-s$ ($\mod 8$), and
the left hand side is well-defined.
We also use the notation
\begin{eqnarray}
KKO^{-n}(X,Y)=KKO_n(C_0(X,\R),C_0(Y,\R))
\end{eqnarray}
when the $C^*$-algebras $\cA$ and $\cB$ are commutative
and associated to topological spaces $X$ and $Y$.
In particular, when $X$ is a point,
they are related to the real K-theory as
\begin{eqnarray}
KKO^{-n}(pt,Y)=KO^{-n}(Y).
\end{eqnarray}

In type I string theory,
as we mentioned in the introduction,
the K-theory group for the unstable D$p$-brane system
wrapped on $Y$, which classifies the charge of D-branes
embedded in $Y$, is $KO^{p-1}(Y)$ \cite{Be}.
The analogous argument as that given for the type II string theory
in section \ref{KKcomplex}
suggest that D-branes embedded in the space $X\times Y$ made from
the unstable D$p$-brane system wrapped on $Y$ are
classified by the KK-theory group $KKO^{p-1}(X,Y)$.
In the next section,
we will give the physical interpretation of the elements
of $KKO^{p-1}(X,Y)$ in terms of the world-volume theory
of the unstable D$p$-brane system, and show more explicitly
that this is the relevant group.

\section{Real KK-theory and type I non-BPS D-branes}
\label{realKKvsD}

Let us compare the real KK-theory groups and
the world-volume theory of non-BPS D$p$-branes.
First,
the elements of $KKO^0(X,Y)$ can be interpreted in
terms of D$p$-brane - anti D$p$-brane system
($p=1,9$) wrapped on $Y$ in type I string theory,
using the exactly analogous argument as in the type II case.

Then, let us consider $KKO^{p-1}(X,Y)$ for $p\ne 1,9$.
{}From (\ref{KKO_n}), we have
\begin{eqnarray}
KKO^{-n}(X,Y)&=&KKO_n(C_0(X,\R),C_0(Y,\R)),\\
&=&
\left\{
\begin{array}{lc}
KKO(C_0(X,\R),C_0(Y,\R)\otimes \C^{n,0})&(n>0),\\
KKO(C_0(X,\R),C_0(Y,\R)\otimes \C^{0,-n})&(n<0).
\label{KKOC}
\end{array}
\right.
\end{eqnarray}
We use (\ref{KKOC})
as the definition of the KK-groups
$KKO^{-n}(X,Y)$ and give a physical interpretation to them.

$KKO(C_0(X,\R),C_0(Y,\R)\otimes \C^{p,q})$
consists of equivalence classes of triples $(\wh\cH,\wh\phi,F)$,
where $\wh\cH$ is the
real Hilbert space over $C_0(Y,\R)\otimes \C^{p,q}$,
$\wh\phi : C_0(X,\R)\ra \B(\wh\cH)$ is a *-homomorphism,
and $F$ is a self-adjoint operator in $\B(\wh\cH)$.
We also require that
$\wh\phi(a)$ ($a\in C_0(X,\R)$) is even 
and $F$ is odd with respect to the $\Z_2$-grading.
The equivalence relations are again given by
unitary equivalence, operator homotopy and
addition of degenerate elements, which are defined in analogous
way as explained in section \ref{KKcomplex} for the complex KK-theory.
Here the unitary equivalence is given by even unitary operators
in $\B(\wh\cH)$.

Since $\wh\cH\simeq\cH\otimes\C^{p,q}$,
where $\cH=C_0(Y,\R)^\infty$, one can show that
$\B(\wh\cH)\simeq \B(\cH)\otimes\C^{p,q}$.
Therefore, the operator $F$ and $\wh\phi(a)$ for
$a\in C_0(X,\R)$ can be expressed as
\begin{eqnarray}
F&=&\sum_{v_n\in\C_{odd}^{p,q}} T_n v_n,
\label{F}\\
\wh\phi(a)&=&\sum_{w_n\in\C_{even}^{p,q}} \Phi_n w_n,
\end{eqnarray}
where $T_n,\Phi_n\in\B(\cH)$.
Here $\C^{p,q}_{even}$ and $\C^{p,q}_{odd}$
denote the sets of even and odd elements in $\C^{p,q}$
spanned by the basis $w_n$ and $v_n$, respectively.

We will show in the following that
the operators $\wh\phi(a)$ ($a\in C_0(X,\R)$)
and $F$ correctly behave as the scalar and tachyon fields
listed in table \ref{unstable}, respectively.
For this purpose, we can restrict our consideration
to the configurations which are constant along $Y$
without any loss of generality.
So, we will set $Y$ to be a point for simplicity.
In this case, $\cH$ is just a separable Hilbert space
over $\R$ and $\B(\cH)$ is the set of bounded linear operators on the
Hilbert space $\cH$.
Thanks to the Bott periodicity
($KKO^n(X,Y)\simeq KKO^{n\pm 8}(X,Y)$), it is sufficient
to consider $KKO^{n}(X,Y)$ with $-3\le n\le 4$.
Therefore, we will consider
$KKO^{-n}(X,pt)=KKO(C_0(X,\R),\C^{n,0})$ ($n=1,2,3$)
and 
$KKO^{n}(X,pt)=KKO(C_0(X,\R),\C^{0,n})$ ($n=1,2,3,4$)
in the following.

\subsection{Non-BPS D0, D8-brane : $KKO^{-1}(X,Y)$}

Let us consider $KKO^{-1}(X,pt)=KKO(C_0(X,\R),\C^{1,0})$.
The generator $e_1$ of $\C^{1,0}$ satisfies $e_1^2=-1$,
and it can be represented by $i$ in $\C$.
Thus $\C^{1,0}=\R\oplus \R i=\C$, 
where real (imaginary) part consists of the even (odd) elements.

The gauge transformation is identified
as the unitary transformation on $\wh\cH$
which is even with respect to the $\Z_2$-grading.
In this case, they are real unitary
operators, which implies
that the gauge group is the orthogonal group $O(\infty)$.

The tachyon is identified as an odd operator $F=iT$,
where $T$ is a real operator,
with the self-adjoint condition $F^\dag=F$.
Therefore, $T$ should be an anti-symmetric operator
($T^T=-T$) which behaves as
the anti-symmetric tensor representation $\asym$
 under the gauge transformation.

The scalar fields are even elements
which means that they are real operators.
Furthermore, they should be self-adjoint since 
they are in the image of the *-homomorphism $\wh\phi$.
Then the scalar fields behave as the symmetric tensor representation
$\sym$ of the gauge group.

These results are consistent
with the world-volume theory of non-BPS D0 or D8-branes.

\subsection{Non-BPS D$(-1)$, D7-brane : $KKO^{-2}(X,Y)$}
\label{Dinst}

Let us consider $KKO^{-2}(X,pt)=KKO(C_0(X,\R),\C^{2,0})$.
The generators of $\C^{2,0}$ are represented
by the elements of the quaternion as $e_1=i$, $e_2=j$,
where $i$ and $j$ are two of the generators of the quaternion
$i,j,k$ which are anti-Hermitian and satisfy 
$i^2=j^2=k^2=-1$, $ij=-ji=k$.
Hence, $\C^{2,0}=\R\oplus \R i\oplus\R j\oplus\R k=\H$,
in which even elements are $\C_{even}^{2,0}=\R\oplus\R k$ and
odd elements are $\C_{odd}^{2,0}=\R i\oplus\R j$.

Since the unitary transformation is given by an even element
which is of the form $g=g_1+g_2 k$, where $g_1$ and $g_2$ are
real operators, satisfying $g^\dag g=1$,
the gauge group consists of (complex) unitary operators
$U(\infty)$.

The tachyon is an odd element $F=T_1i-T_2 j=i(T_1+T_2k)$
satisfying $F^\dag=F$,
where $T_1$ and $T_2$ are real operators.
Let us define $T\equiv T_1+T_2k$,
then $T$ is a complex anti-symmetric operator satisfying $T^T=-T$.
The transformation of tachyon under the gauge group is given by
\begin{eqnarray}
F=iT\ra g^\dag F g=i g^TT g,
\end{eqnarray}
{}from which we can see that
 $T$ is transformed as the anti-symmetric tensor representation
$\asym$, as expected.

The scalar fields are even self-adjoint operators.
This means that the scalar fields are
complex Hermite operators, and they belong to
the adjoint representation of the gauge group $U(\infty)$.

These results are consistent
with the world-volume theory of non-BPS D$(-1)$ or D7-branes.

\subsection{Non-BPS D6-brane : $KKO^{-3}(X,Y)$}
\label{C30}

Let us consider $KKO^{-3}(X,pt)=KKO(C_0(X,\R),\C^{3,0})$.
The generators of $\C^{3,0}$ are faithfully represented as
\begin{eqnarray}
e_1=\mat{i,,,-i}, e_2=\mat{j,,,-j}, e_3=\mat{k,,,-k},
\end{eqnarray}
where $i,j,k$ are the generators of quaternion.
Even elements of $\C^{3,0}$ are of the form
\begin{eqnarray}
\mat{a+bi+cj+dk,,,a+bi+cj+dk}=\mat{a',,,a'},
\end{eqnarray}
and odd elements are
\begin{eqnarray}
\mat{a+bi+cj+dk,,,-a-bi-cj-dk}=\mat{a',,,-a'},
\end{eqnarray}
where $a,b,c,d\in\R$
and $a'=a+bi+cj+dk\in\H$ is a quaternion.

The gauge transformation is given by an even element
\begin{eqnarray}
g=\mat{g',,,g'},
\label{sp}
\end{eqnarray}
with $g^\dag g=1$,
where $g'=g_a+g_b i+g_c j+g_d k$ is
a quaternionic operator.
It is easy to see that the quaternionic unitary operator $g'$
is an element of $Sp$ group. (See Appendix \ref{App2}.)
Thus the gauge group is $Sp(\infty)$ as expected.

The tachyon operator is a self-adjoint odd element of the Clifford algebra,
and hence we can set
\begin{eqnarray}
F=\mat{T,,,-T},~~~T^\dag=T
\end{eqnarray}
where $T=T_0+T_1 i+T_2 j+T_3 k$ is a quaternionic operator.
As shown in Appendix \ref{App2}, this behaves
as the anti-symmetric tensor representation $\asym$ of the gauge group.

The scalar fields are even self-adjoint elements,
\begin{eqnarray}
\Phi=\mat{\Phi',,,\Phi'},~~~{\Phi'}^\dag = \Phi',
\end{eqnarray}
where $\Phi'$ is a quaternionic operator.
Thus they also transform as
the anti-symmetric tensor representation $\asym$ under the gauge
transformation.

All these results are consistent
with the world-volume theory of non-BPS D6-branes.

\subsection{D5 - anti D5 system : $KKO^{-4}(X,Y)$}

Let us consider $KKO^{-4}(X,pt)=KKO(C_0(X,\R),\C^{4,0})$.
The generators for the Clifford algebra $\C^{4,0}$ are
represented as
\begin{eqnarray}
e_1=\mat{,i,i,},e_2=\mat{,j,j,},e_3=\mat{,k,k,},e_4=\mat{,-1,1,}
\end{eqnarray}
where $i,j,k$ are the generators of quaternion.
Then, the even elements are of the form
\begin{eqnarray}
\mat{a+bi+cj+dk,,,e+fi+gj+hk}
\end{eqnarray}
and odd elements are
\begin{eqnarray}
\mat{,a+bi+cj+dk,e+fi+gj+hk,}
\end{eqnarray}
where $a,b,c,d,e,f,g,h\in\R$.

The gauge group consists of even unitary operators,
which is of the form
\begin{eqnarray}
g&=&
\mat{g_1,,,g_2},
\end{eqnarray}
where $g_1$ and $g_2$ are quaternionic operators
satisfying $g_1^\dag g_1=g_2^\dag g_2=1$.
As explained in Appendix \ref{App2},
the unitary quaternionic operators $g_1$ and $g_2$ are
elements of $Sp$ group.
Therefore, the gauge group is $Sp(\infty)\times Sp(\infty)$,

The tachyon operator is a self-adjoint odd operator, which is of the form
\begin{eqnarray}
F=\mat{,T^\dag,T,},
\label{FTT}
\end{eqnarray}
where $T$ is a quaternionic operator.
It transforms as
\begin{eqnarray}
F\ra g^\dag F g =\mat{,g_1^\dag T^\dag g_2,g_2^\dag T g_1,},
\end{eqnarray}
under the gauge transformation,
which show that the operator $T$ transform as
the bi-fundamental representation of the gauge group
$Sp(\infty)\times Sp(\infty)$.

The scalar fields $\Phi$ are even operators of the form
\begin{eqnarray}
\Phi=\mat{\Phi_1,,,\Phi_2},
\end{eqnarray}
where $\Phi_1$ and $\Phi_2$ are self-adjoint quaternionic
operators, which means that $\Phi_i$ ($i=1,2$) belong
to anti-symmetric tensor representation $\asym$ of the gauge group.
(See Appendix \ref{App2}.)

These results are consistent
with the world-volume theory of D5-brane - anti D5-brane
system.

\subsection{Non-BPS D2-brane : $KKO^1(X,Y)$}

Let us consider $KKO^1(X,pt)=KKO(C_0(X,\R),\C^{0,1})$.
The generator $e_1$ of $\C^{0,1}$ is represented by
$e_1=\sigma_3$.
The even element is of the form
\begin{eqnarray}
\mat{a,,,a},
\end{eqnarray}
and the odd element is 
\begin{eqnarray}
\mat{a,,,-a}
\end{eqnarray}
for $a\in\R$.

The gauge group consists of even elements, and hence
 real unitary operators, namely $O(\infty)$.

The tachyon is an odd element which is of the form
$F=T\sigma_3$, where $T$ is a real operator,
with $F^\dag=F$, which means that $T$ is
the symmetric tensor representation $\sym$ of the gauge group.

The scalar fields are even self-adjoint elements,
which also belong to the symmetric tensor representation
 $\sym$ of the gauge group $O(\infty)$.

These results are consistent with the world-volume theory of
non-BPS D2-branes.

\subsection{Non-BPS D3-brane : $KKO^2(X,Y)$}

Let us consider $KKO^2(X,pt)=KKO(C_0(X,\R),\C^{0,2})$.
The generators of $\C^{0,2}$ are represented
by $e_1=\sigma_1$ and $e_2=\sigma_2$.
An even element is of the form
\begin{eqnarray}
\mat{a+ib,,,a-ib},
\end{eqnarray}
and an odd element is
\begin{eqnarray}
\mat{,a-ib,a+ib,},
\end{eqnarray}
where $a,b\in\R$.

The gauge group consists of even elements of the form
\begin{eqnarray}
g=\mat{g_a+ig_b,,,g_a-ig_b}\equiv\mat{g',,,\ol g'}.
\end{eqnarray}
Therefore the gauge group is $U(\infty)$.

The tachyon operator is an odd element
\begin{eqnarray}
F=\mat{,\ol T,T,}
\end{eqnarray}
where $T=T_a+T_b i$ is a complex operator.
The self-adjoint condition $F^\dag=F$ implies
$T=T^T$, which means that
the tachyon belongs to the symmetric tensor representation $\sym$
of the gauge group.
In fact, the gauge transformation $F\ra g^\dag F g$
becomes
\begin{eqnarray}
T\ra {g'}^T T g',
\end{eqnarray}
which is the correct transformation of the
symmetric tensor representation $\sym$.

The scalars are even self-adjoint elements,
and hence they belong to the adjoint representation of the
gauge group $U(\infty)$.

These results are consistent with the world-volume theory
of non-BPS D3-branes.

\subsection{Non-BPS D4-brane : $KKO^3(X,Y)$}

\label{C03}

Let us consider $KKO^3(X,pt)=KKO(C_0(X,\R),\C^{0,3})$.
The generators of $\C^{0,3}$ are represented as
$e_1=\sigma_1$, $e_2=\sigma_2$ and $e_3=\sigma_3$.
The even elements are
\begin{eqnarray}
a+b i\sigma_1+ c i\sigma_2+ d i\sigma_3,
\end{eqnarray}
and odd elements are
\begin{eqnarray}
a i+b \sigma_1+ c \sigma_2+ d \sigma_3,
\end{eqnarray}
where $a,b,c,d\in\R$.

The gauge transformation is given by even unitary operator,
which is of the form
\begin{eqnarray}
g&=&g_a+g_b i\sigma_1+ g_c i\sigma_2+ g_d i\sigma_3\\
&=&\mat{g_a+ig_d,g_c+ig_b,-g_c+ig_b,g_a-ig_d },
\end{eqnarray}
satisfying $g^\dag g=1$.
This is nothing but the $Sp$ group. (See Appendix \ref{App2}.)

It may be convenient to represent $i\sigma_i$,
by the generators of quaternion as
$i\sigma_1\ra i$, $i\sigma_2\ra j$ and $i\sigma_3\ra -k$.
Then the gauge group is represented as
quaternionic unitary operator, which we encountered in (\ref{sp}).

Then, the tachyon operator is an odd element
$F=iT_0+T_1\sigma+T_2\sigma_2+T_3\sigma_3\equiv iT$ satisfying
$F^\dag=F$. If we express the operator $T$
in the quaternionic representation,
we see that
\begin{eqnarray}
T=T_0-T_1 i-T_2 j+T_3 k,~~~T^\dag=-T,
\label{sym2}
\end{eqnarray}
which transforms as
\begin{eqnarray}
T\ra g^\dag T g,
\end{eqnarray}
under the gauge transformation.
Therefore, as shown in Appendix \ref{App2},
the tachyon is transformed as the adjoint representation ($\sym$)
of the gauge group $Sp(\infty)$.

The scalars are even elements.
Using the quaternionic representation, it can be written as
\begin{eqnarray}
\Phi=\Phi_a+\Phi_bi+\Phi_cj+\Phi_dk,~~\Phi^\dag=\Phi,
\end{eqnarray}
which transform as the anti-symmetric tensor representation
$\asym$ of the gauge group,
as shown in Appendix \ref{App2}.

These results are consistent with the
world-volume theory of non-BPS D4-branes.

\section{An explicit construction of flat D$p$-branes}
\label{flat}

Let us make things more explicit in some simple situations.
In this section, we consider flat D-branes in flat space-time,
mainly using the matrix theory based on
an infinite number of non-BPS D-instantons in type I string theory,
which we call type I K-matrix theory.

\subsection{ Flat D$p$-branes in type I K-matrix theory}
The type I K-matrix theory resembles IIB matrix theory
\cite{IKKT} in many respects.
In fact, it is a matrix theory with unitary gauge symmetry,
and the field contents include scalar fields and fermions
which constitute vector and spinor representation
of the Lorentz group $SO(1,9)$, respectively, and
transform as the adjoint representation of the gauge group.
But there are some important differences.
First, we have some extra matters including a tachyon,
 and then, we have to take the size of the matrices to be
infinity from the beginning in order that we can create
an arbitrary number of non-BPS D-instantons.

As we have argued, D-branes in the type I K-matrix theory
are classified by $KKO^{-2}(X,pt)$.
Let us consider flat D$p$-branes extended in the
$x^0,\dots,x^p$ directions in flat space-time $\R^{10}$
in the type I K-matrix theory.
These configurations are classified by $KKO^{-2}(\R^{p+1},pt)$.
\footnote{
Here $\R^{p+1}$ is not compact, but
 $KKO^{-2}(\R^{p+1},pt)=KKO_2(C_0(\R^{p+1},\R),\R)$ is well-defined.
It satisfies
$KKO^{-2}(S^{p+1},pt)=KKO^{-2}(\R^{p+1},pt)\oplus KKO^{-2}(pt,pt)$,
where $S^{p+1}$ is the one point compactification of $\R^{p+1}$.
The left hand side $KKO^{-2}(S^{p+1},pt)$ classifies D-branes
on $S^{p+1}$, and $KKO^{-2}(pt,pt)$ in the right hand side
 classifies D-instantons sitting at a point
in the space-time.
The only possible
stable D-branes on $S^{p+1}$ are  D$p$-branes
wrapped on the $S^{p+1}$ and the D-instantons. Thus,
we interpret $KKO^{-2}(\R^{p+1},pt)$ as the group which
classifies the D$p$-branes.
}

Let us first consider the $p=-1$ case.
The group $KKO^{-2}(pt,pt)$ consists of homotopy classes
of the anti-symmetric operator $T$ acting on a Hilbert space $\cH$
(over $\C$), satisfying $T^\dag T-1 \in \K(\cH)$.
This condition implies that $\ker T$ is a finite dimensional
vector space. Since $T$ is anti-symmetric,
$\dim \ker T$ $(\mod 2)$ is invariant under small perturbation
of the operator $T$.
In fact, we have $KKO^{-2}(pt,pt)=KO^{-2}(pt)=\Z_2$.
Therefore, a single non-BPS D-instanton is stable, while
a pair of non-BPS D-instantons can be annihilated,
in agreement with the results in \cite{Wi}.

For generic $p$,
we obtain
\begin{eqnarray}
KKO^{-2}(\R^{p+1},pt)=KO(\R^{9-p})=
\left\{
\begin{array}{cl}
\Z&(p=1,5,9~~(\mod 8))\\
\Z_2&(p=-1,0,7,8~~(\mod 8))\\
0&(\mbox{others})
\end{array}
\right.
\label{Dspec}
\end{eqnarray}
 using the isomorphism
\begin{eqnarray}
KKO^{k}(X,Y)=KKO^{k-n}(X\times \R^n,Y)
=KKO^{k+m}(X,Y\times \R^m).
\label{KKisom}
\end{eqnarray}
Hence we have $\Z_2$ charge D$p$-branes ($p=-1,0,7,8$)
and $\Z$ charge D$p$-branes ($p=1,5,9$), which is exactly
what we expect from the K-theory analysis \cite{Wi}, i.e.
table \ref{spec}.

The explicit configuration representing D$p$-brane can be obtained
by finding the configuration in type IIB K-matrix theory
(i.e. the matrix theory based on D-instanton - anti D-instanton system 
in type IIB theory)
which survive after projecting out the unwanted components of the matrices
in type I theory.

Recall that there are
ten pairs of scalars $\Phi^\mu$, ${\ol\Phi}^\mu$ ($\mu=0,1,\dots,9$),
which represent the position of D-instantons and anti D-instanton
respectively, together with a tachyon $T$ in the IIB K-matrix theory.
A configuration representing a D$p$-brane
in the IIB K-matrix theory \cite{Te,AsSuTe} is given by
\begin{eqnarray}
&&T=u\sum_{\alpha=0}^p \wh p_\alpha\otimes \gamma^\alpha,
\label{T_IIB}
\\
&&\Phi^\alpha=\ol{\Phi}^{\alpha}=\wh{x}^{\alpha} \otimes {\rm 1}
~~~~~(\alpha=0, \cdots, p), \\
&&\Phi^i=\ol{\Phi}^i=0
~~~~~(i=p+1, \cdots, 9),
\label{Phi_IIB}
\end{eqnarray}
which act on
the Hilbert space $L^2(\R^{p+1})\otimes S$, where $S$ is the vector
space on which the matrices $\gamma^\alpha$ are represented.
Here $\wh x^\alpha$ is defined by multiplication of $x^\alpha$
and $\wh p_\alpha=-i\del_\alpha=-i\del/\del x^\alpha$ is a differential
operator, both acting on $L^2(\R^{p+1})$, and
\begin{eqnarray}
\Gamma^\alpha\equiv
\mat{,{\gamma^\alpha}^\dag,\gamma^\alpha,}
\label{Gamma}
\end{eqnarray}
are $SO(p+1)$ gamma matrices
satisfying $\{\Gamma^\alpha,\Gamma^\beta\}=2\delta^{\alpha\beta}$.
(\ref{T_IIB}) can also be written as
\begin{eqnarray}
F\equiv\mat{,T^\dag,T,}=u\sum_{\alpha=0}^p\wh p_\alpha\otimes\Gamma^\alpha
=u\sum_{\alpha=0}^p \del_\alpha\otimes\wh\Gamma^\alpha,
\label{FDp}
\end{eqnarray}
where we have defined $\wh\Gamma^\alpha=-i\Gamma^\alpha$.
Note that if $p$ is even, we can choose $\gamma^\alpha$ to be
Hermitian, which form the irreducible $SO(p+1)$ gamma matrices.
The size of the matrices $\gamma^\alpha$ are listed
in table \ref{IIBgamma}.
\begin{table}[htb]
\begin{center}
$$
\begin{array}{c|cccccccccc}
\hline\hline
p&0&1&2&3&4&5&6&7&8&9\\
\hline
\mbox{size}&1&1&2&2&4&4&8&8&16&16\\
\hline
\end{array}
$$
\parbox{75ex}{
\caption{
\small
The size of the matrices $\gamma^\alpha$ in (\ref{T_IIB})
used for the tachyon configuration
representing type IIB D$p$-branes.
}
\label{IIBgamma}
}
\end{center}
\end{table}
It can be shown that this configuration becomes an exact D$p$-brane solution
of the BSFT action for type IIB D-instanton - anti D-instanton
system \cite{KuMaMo,KrLa,TaTeUe},
if we take $u\rightarrow \infty$ \cite{Te}.
\footnote{
Note that the minimum of the tachyon potential of the BSFT action
is given by $TT^\dag=T^\dag T=\infty\cdot 1$. This is the reason that
we chose an unbounded operator for the tachyon in (\ref{T_IIB}).
If one wish to normalize the tachyon
so that the minimum is given by $T'{T'}^\dag={T'}^\dag T'=1$,
as in section \ref{KKcomplex},
one can redefine the tachyon as $T'=T/\sqrt{1+T^\dag T}$.
}
The solution represents a BPS D$p$-brane for odd $p$ and a
non-BPS D$p$-brane for even $p$.
The tension and RR-charge
for the solution can be calculated in the BSFT framework
and it is also shown in \cite{Te} that
they exactly reproduce the expected value.

The part of the action of the type I non-BPS D-instantons which include
only $\Phi^\mu$ and $T$ is obtained by projecting out the unwanted
components of the matrices $\Phi^\mu$ $\ol{\Phi}^\mu$ and $T$ from the 
type IIB BSFT action,
at least at the tree level.
Therefore,
the solution of type IIB K-matrix theory satisfying
$\Phi^\mu=\ol{\Phi}^\mu$ and $T^T=-T$ is automatically
a solution of type I K-matrix theory.
In order for $T$ to be anti-symmetric,
$\gamma^\alpha$ should be symmetric matrices,
since the differential operators $\wh p_\alpha$ are anti-symmetric.
It is easy to find a representation of $SO(p+1)$ gamma matrices
$\Gamma^\alpha$ of the form (\ref{Gamma}) with symmetric $\gamma^\alpha$.
First, note that one can always set $\Gamma^0$ as
\begin{eqnarray}
\Gamma^0&=&1\otimes\sigma^2,
\label{gamma0}
\end{eqnarray}
using unitary transformation.
Then, the condition $\{\Gamma^\alpha,\Gamma^\beta\}=2\delta^{\alpha\beta}$
implies
\begin{eqnarray}
\Gamma^i&=&\gamma^i_p\otimes\sigma^1~~~(i=1,\dots,p),
\label{gammai}
\end{eqnarray}
where $\gamma_p^i$ are $SO(p)$ gamma matrices represented as
real symmetric matrices.
Note also that we can identify $-i\sigma^2$ and $-i\sigma^1$ as
the generators $e_1$ and $e_2$ of the Clifford algebra $\C^{2,0}$,
and the operator $F$ in (\ref{FDp}) can be written of the form
\begin{eqnarray}
F=\mat{,T^\dag,T,}=u\left(\del_0\, e_1+\del_i\gamma_p^i\, e_2\right).
\label{Dpsol}
\end{eqnarray}

The minimum size of the matrices $\gamma^\alpha$
are listed in table \ref{Igamma},
which can be obtained from table \ref{Clifford} in appendix \ref{App1},
since the $SO(p)$ gamma matrices $\gamma^i_p$ form a real representation
of $\C^{0,p}$.
\begin{table}[htb]
\begin{center}
$$
\begin{array}{c|cccccccccc}
\hline\hline
p&0&1&2&3&4&5&6&7&8&9\\
\hline
\mbox{size}&1&1&2&4&8&8&16&16&16&16\\
\hline
\end{array}
$$
\parbox{75ex}{
\caption{
\small
The size of the matrices $\gamma^\alpha$
 for the tachyon configuration
representing type I D$p$-branes.
}
\label{Igamma}
}
\end{center}
\end{table}
Comparing table \ref{IIBgamma} and table \ref{Igamma},
we see that the size of the tachyon in type I
with $p=3,4,5,6,7$
is twice the size in type IIB.
This implies that
the tension of a D$p$-brane in type I theory
are twice that in type IIB theory
for $p=3,4,5,6,7$, while they are the same for $p=0,1,2,8,9$,
up to the common factor $1/\sq$.
This result is consistent with the construction of D-branes
in type I theory as given in \cite{Wi,Be}. Namely, a D$p$-brane
in type I theory is given by the unoriented projection
of a single D$p$-brane ($p=0,1,2,8,9$), 
two D$p$-branes ($p=4,5,6$) or a D$p$-brane - anti D$p$-brane pair
($p=-1,3,7$) in type IIB theory, as one can read from
the gauge group listed in table \ref{unstable}.

Let us next consider the tachyonic mode around the
D$p$-brane solution (\ref{Dpsol}).
Suppose that the tachyon $F$ is the sum $F_0+F'$ of
the solution $F_0$ and the fluctuation $F'$ satisfying
\begin{eqnarray}
\{F_0,F'\}=0,~~~{F'}^\dag=F'.
\label{fluc}
\end{eqnarray}
Then, $F^2=F_0^2+{F'}^2$ implies that $F'$ has negative mass squared,
which means that $F'$ is the tachyonic mode on the D$p$-brane.
In fact, inserting (\ref{fluc}) into the BSFT action for the non-BPS
D-instanton system, one can easily see that the potential for the
fluctuation $F'$ is exactly what we expect for the D$p$-brane.
Such fluctuation around the solution (\ref{Dpsol}) is of the form
\begin{eqnarray}
F'=T'\wh\gamma\,e_2,
\end{eqnarray}
where $T'$ is a real parameter
and $\wh\gamma$ is a real matrix, whose size is the
same as that of $\gamma_p^i$ listed in table \ref{Igamma}, satisfying
\begin{eqnarray}
\{\wh\gamma,\gamma_p^i\}=0,~~~\wh\gamma^\dag=-\wh\gamma,~~~\wh\gamma^2=-1.
\end{eqnarray}
Note that $\gamma_p^i$ ($i=1,\dots,p$) together with $\wh\gamma$
make up a real representation of $\C^{1,p}$.
In order that such $\wh\gamma$ exists, there must be
a real representation of $\C^{1,p}$ whose dimension
is the same as the size of $\gamma_p^i$.
The dimension of the irreducible real representation of $\C^{1,p}$
is listed in table \ref{C1p}, which is
obtained using the identity $\C^{1,p}=M_2(\C^{0,p-1})$ (See (\ref{Cisom}).)
and table \ref{Clifford}.
\begin{table}[htb]
\begin{center}
$$
\begin{array}{c|cccccccccc}
\hline\hline
p&0&1&2&3&4&5&6&7&8&9\\
\hline
\mbox{dim.}&2&2&2&4&8&16&16&32&32&32\\
\hline
\end{array}
$$
\parbox{75ex}{
\caption{
\small
The dimension of the irreducible real representation
of $\C^{1,p}$.
}
\label{C1p}
}
\end{center}
\end{table}
Therefore, comparing table \ref{Igamma} and table \ref{C1p},
we conclude that there is a tachyonic mode for
$p=2,3,4,6$, which agrees with the standard
result first derived in \cite{Wi}.

Then, what happens if there are two D$p$-branes for
$p=0,1,5,7,8,9$?
The solution representing two D$p$-branes is obtained
by simply tensoring the rank two unit matrix $1_2$ with
the solution (\ref{Dpsol}) as
\begin{eqnarray}
F=u\,1_2\otimes
\left(\del_0\, e_1+\del_i\gamma_p^i\, e_2\right).
\end{eqnarray}
In this case, the fluctuation
\begin{eqnarray}
F'=T'\,\epsilon\otimes \gamma^{p+1} e_2
\end{eqnarray}
satisfies the condition (\ref{fluc}).
Here $\epsilon=i\sigma_2$ and $\gamma^{p+1}$ is a real matrix,
 whose size is the same as that of $\gamma_p^i$,
 satisfying
\begin{eqnarray}
\{\gamma^{p+1},\gamma_p^i\}=0,~~~\gamma^{p+1\dag}=\gamma^{p+1},
~~~(\gamma^{p+1})^2=1.
\end{eqnarray}
Namely, $\gamma_p^i$ ($i=1,\dots,p$) together with $\gamma^{p+1}$
make up a real representation of $\C^{0,p+1}$.
The analogous argument as above implies that
there is a tachyonic mode if the dimension of the real representation
of $\C^{0,p+1}$, which is given by shifting $p$ to $p+1$ in
table \ref{Igamma}, is equal to the size of the matrices $\gamma_p^i$
listed in table \ref{Igamma}. Hence, we conclude
that two coincident
non-BPS D$p$-branes with $p=0,7,8$ have tachyon modes,
which again reproduces the results in \cite{Wi}.
All these results are consistent with the spectrum of D-branes
in type I theory (table \ref{spec}).

\subsection{ Flat D$p$-branes in the unstable D$q$-brane system}
Now we generalize the above consideration and
describe the construction of flat D$p$-branes from
unstable D$q$-brane system with $q\ge 0$ in type I theory.
Let us consider the unstable D$q$-brane system extended in the
$x^0,x^1,\dots,x^q$ directions, and construct
D$p$-branes along $x^0,\dots,x^{q-m}$ and $x^{q+1},\dots,x^{p+m}$
directions.
Generalizing the argument above,
we expect that such D$p$-branes are classified by
$KKO^{q-1}(\R^{p+m-q},\R^m)$.
Using the isomorphism (\ref{KKisom}), we can again show
\begin{eqnarray}
KKO^{q-1}(\R^{p+m-q},\R^m)=KO(\R^{9-p})
\end{eqnarray}
which reproduces the correct spectrum of the type I D-branes
as in (\ref{Dspec}).

The explicit tachyon configuration is given by
\begin{eqnarray}
F=u\sum_{\alpha=q-m+1}^q \wh x_\alpha\otimes\Gamma^\alpha
+u\sum_{\beta=q+1}^{p+m}\del_\beta\otimes\wh\Gamma^\beta,
\end{eqnarray}
where
$\Gamma^\alpha$ and $\wh\Gamma^\beta$ are elements of
$M_n(\R)\otimes \C^{1,q}_{odd}$ for some $n$,
\footnote{
Here we use $\C^{1,q}$ for the definition
of the KK-theory group as
$KKO^{q-1}(X,Y)=KKO(C_0(X,\R),C_0(Y,\R)\otimes\C^{1,q})$.
}
satisfying
\begin{eqnarray}
&&\Gamma^{\alpha\dag}=\Gamma^\alpha,
~~~\wh\Gamma^{\beta\dag}=-\wh\Gamma^\beta,\\
&&\{\Gamma^\alpha,\Gamma^{\alpha'}\}=2\delta^{\alpha\alpha'},~~~
\{\wh\Gamma^\beta,\wh\Gamma^{\beta'}\}=-2\delta^{\beta\beta'},~~~
\{\Gamma^\alpha,\wh\Gamma^{\beta}\}=0,
\end{eqnarray}
which are the same as the relations for the generators of $\C^{p+m-q,m}$.

A realization of these $\Gamma^\alpha$
and $\wh\Gamma^\beta$
representing a single D$p$-brane is given as follows.
Let us first consider $m=q$ case. One can find the realization for
this case
generalizing the consideration around (\ref{gamma0}) and (\ref{gammai});
\begin{eqnarray}
\wh\Gamma^{q+i}&=&\gamma_p^i\otimes e_1,~~~(i=1,\dots,p),
\label{gamma1}\\
\Gamma^{i}&=&1\otimes e_{i+1},~~~(i=1,\dots,q),
\label{gamma2}
\end{eqnarray}
where $e_i$ ($i=1,\dots,q+1$) are the generators
of $\C^{1,q}$.
They are realized as elements of $M_{n_p}(\R)\otimes \C^{1,q}_{odd}$,
where $n_p$ is the size of the $SO(p)$ gamma matrices $\gamma_p^i$
listed in table \ref{Igamma}.

Then, let us consider general cases with $m\le q$.
Note that using the isomorphism
 $\C^{p,q}=M_{2^{q-m}}(\R)\otimes\C^{p+m-q,m}$
(See (\ref{Cisom}).),
we can embed $\C^{p+m-q,m}_{odd}$ as
 $\diag (1,0,\dots,0)\otimes\C^{p+m-q,m}_{odd}$ in
$\C^{p,q}_{odd}$.
Thus we can realize $\C^{p+m-q,m}_{odd}$ in $M_n(\R)\otimes\C_{odd}^{1,q}$
using the realization of $\C^{p,q}$ given by
(\ref{gamma1}) and (\ref{gamma2}).

The corresponding configurations in type IIB theory
are given by replacing the real gamma matrices $\gamma_p^i$ with
complex gamma matrices.
In any cases, difference of the size of the tachyon
between type I and  type IIB is again determined by
comparing table \ref{Igamma} and table \ref{IIBgamma},
which implies the correct tension for the D$p$-brane.

As a check, let us demonstrate the case with $q=9$ and $m=9-p$.
In this case, the tachyon is given by
\begin{eqnarray}
F=u\sum_{\alpha=p+1}^9 \wh x_\alpha\otimes\Gamma^\alpha,
\label{D9case}
\end{eqnarray}
where $\Gamma^\alpha$ ($\alpha=p+1,\dots,9$)
are elements of $M_n(\R)\otimes \C^{1,9}_{odd}$
satisfying the same relations as that for the
generators of $\C^{0,9-p}$.
As explained above, $\C^{0,9-p}$ can be embedded in
$\C^{p,9}=M_{2^p}(\R)\otimes\C^{0,9-p}$ as a subalgebra
of the form $\diag (1,0,\dots,0)\otimes \C^{0,9-p}$,
and the generators of $\C^{p,9}$ can be realized in
$M_{n_p}(\R)\otimes\C^{1,9}_{odd}$ by (\ref{gamma1}) and (\ref{gamma2}).
Note that the relations (\ref{Cisom}) and (\ref{Bott}) imply
 $M_{n_p}(\R)\otimes\C^{1,9}=M_{32n_p}(\R)$.
Combining these together, we obtain a realization
of $\Gamma^\alpha$ in $M_{32n_p/2^p}(\R)$.

On the other hand, the tachyon configuration
representing D$p$-brane in D9-brane - anti D9-brane system
is given in \cite{Wi} as (\ref{D9case}) with
real gamma matrices $\Gamma^\alpha$ of the form
(\ref{Gamma}). The matrices $\Gamma^\alpha$ together with
$\Gamma\equiv \diag(1,-1)$ form a real irreducible
representation of $SO(10-p)$ gamma matrices.
Therefore the size of these matrices is
given by $n_{10-p}$, where $n_p$ is the size of
the real $SO(p)$ gamma matrices listed in table \ref{Igamma}.
We can check that $n_{10-p}$ is equal to $32n_p/2^p$,
in agreement with the above consideration.
We can also calculate the ratio of the
tension of a type I D$p$-brane to that of
a type IIB D$p$-brane, comparing the size of the real 
and complex  $SO(10-p)$ gamma matrices, which again
gives the correct values.

\section{Conclusion and Discussions}
\label{Disc}
In this paper, we have examined D-branes in type I string theory.
Our main claim is that the D-branes in $X\times Y$
made from the unstable D$p$-brane system wrapped on $Y$
are classified by real KK-theory groups $KKO^{p-1}(X,Y)$.
We have explicitly shown that the elements of $KKO^{p-1}(X,Y)$
can be interpreted in terms of the field theory
on the unstable D$p$-brane system.
It is quite interesting that 
once we accept the physical interpretation
of $KKO^{p-1}(X,Y)$,
we can easily find the field content
of type I unstable D$p$-brane systems listed in table \ref{unstable},
which was derived by careful consideration
of the $\Omega$-projection in \cite{Be},
by just looking at the Clifford algebra.

The arguments in this paper are also applicable to the $USp(32)$
string theory \cite{Su}.
The D9-brane - anti D9-brane system of the theory has
gauge group of $Sp\times Sp$ type, and the relevant K-theory group
is $KSp(X)$ which is isomorphic to $KO^4(X)$.
Therefore,
the spectrum of the D-branes are obtained by shifting $p$ to $p+4$
in table \ref{spec}, which implies that
there are stable D$1,5,9$-branes
with $\Z$ charges and D$3,4$-branes with $\Z_2$ charges.
Accordingly, the KK-theory group which corresponds
to the unstable D$p$-brane system in this theory is $KKO^{p+3}(X,Y)$.
The field contents of the unstable D$p$-brane system
should also be obtained by the shift $p\ra p+4$ in table \ref{unstable}.

It is not fully clear to us why such Clifford algebra structure
appears in the type I D-branes.
As discussed around (\ref{KKO_n}),
we can use the Clifford algebra $\C^{1,p}$ to define
$KKO^{p-1}(X,Y)$ as
$KKO^{p-1}(X,Y)=KKO(C_0(X,\R),C_0(Y,\R)\otimes\C^{1,p})$.
One can imagine that the algebra $\C^{1,p}$ has to do with
the fermions on the D$p$-brane which transform as a spinor
representation of $SO(1,p)$. Actually, the gauge groups of the
systems are derived in \cite{Be} so that we can consistently
perform the $\Omega$-projection for the fermions created by
the open strings connecting the D$p$-brane and one of the
background D9-branes. It would be interesting to study the
relationship between the Clifford algebras and orientifolds.
See also \cite{Gu} for the related discussions.

Another interesting application of the Clifford algebra
in type I D-branes is that we can write down
the Chern-Simons terms using the superconnection
which are defined by the Clifford algebra.
Generalizing the definition of the (complex) superconnection in
\cite{Qu}, we can define essentially eight types of real superconnections
associated to the corresponding Clifford algebra.
Namely, the real superconnection
associated to the Clifford algebra $\C^{p,q}$ is given as
the formal sum of the tachyon, which is an
$M_n(\R)\otimes \C^{p,q}_{odd}$ valued field, and the gauge field,
which is an $M_n(\R)\otimes \C^{p,q}_{even}$ valued one-form.
More explicitly, we can write it as
\begin{eqnarray}
\cA=\sum_{v_n\in\C_{odd}^{p,q}}T_n v_n+
\sum_{w_n\in\C_{even}^{p,q}}A_n w_n,
\end{eqnarray}
where $T_n$ are the $M_n(\R)$ valued fields and
$A_n$ are the $M_n(\R)$ valued one-form fields, and
 $\{w_n\}$ and $\{v_n\}$
are the basis of $\C^{p,q}_{even}$ and $\C^{p,q}_{odd}$, respectively.
The first term in the right hand side is nothing but
the tachyon $F$ in (\ref{F}), though we have described in a finite rank
matrices here. The first term is required to be self-adjoint,
as discussed in section \ref{realKKvsD},
while the second term should be anti self-adjoint,
since the gauge fields are associated to the Lie algebra of the gauge group.
The Chern-Simons term is of the form
\begin{eqnarray}
S_{CS}=\int C\wedge\mbox{Str}\, e^{\cF},
\end{eqnarray}
where $C$ is the sum of RR-fields,
 $\cF=d\cA+\cA^2$ is the field strength and
`Str' denote the trace of the coefficient of
$e_1e_2\cdots e_{p+q}$.

The most simple and interesting choice of the
unstable D$p$-brane systems is the lowest dimensional
case $p=-1$, i.e.
the matrix theory based on non-BPS D-instantons, which we called
K-matrix theory.
As a check, we have shown that the spectrum of the
flat D-branes constructed in the K-matrix theory
are exactly what we expect in the K-theory
result \cite{Wi}.
This suggests that we can construct
any configurations of D-branes in type I theory
from an infinite number of non-BPS D-instantons.
Therefore, it is natural to suppose that we can study
dynamics involving various types of D-branes within a single
framework of the K-matrix theory, which was one of
the main motivation for our previous paper \cite{AsSuTe}.
It would be interesting to explore further in this direction.

\section*{Acknowledgments}

\vskip2mm
This work was supported in part by JSPS Research Fellowships for Young 
Scientists.

\appendix
\setcounter{equation}{0}
\section{Clifford Algebra}
\label{App1}
The Clifford algebra $\C^{p,q}$ is defined as an algebra over $\R$
generated by $e_i$ ($i=1,\dots,p+q$) satisfying
\begin{eqnarray}
&e_ie_j+e_je_i=0& (i\ne j)
\label{rel1}\\
&e_i^2=-1,~~e_i^*=-e_i& (i=1,\dots,p)
\label{rel2}\\
&e_i^2=1,~~e_i^*=e_i& (i=p+1,\dots,p+q).
\label{rel3}
\end{eqnarray}
It is a $\Z_2$-graded algebra, defined by the involution
$e_i\ra -e_i$. And, $\C_{even}^{p,q}$ and $\C_{odd}^{p,q}$
denote the sets of even and odd elements
with respect to this gradation, respectively.

Here we list some useful isomorphisms among the Clifford algebras.
\begin{eqnarray}
\C^{r+n,s+n}&\simeq& M_{2^n}(\C^{r,s})
\label{Cisom}
\end{eqnarray}
\begin{eqnarray}
\C^{r,s}&\simeq&\left\{
\begin{array}{cc}
M_{2^s}(\C^{r-s,0})&(r>s)\\
M_{2^r}(\C^{0,s-r})&(r<s)
\end{array}
\right.
\end{eqnarray}
\begin{eqnarray}
\C^{r+4,s}\simeq\C^{r,s+4},~~
\C^{r,s+1}\simeq\C^{s,r+1},
\end{eqnarray}
\begin{eqnarray}
\C^{r+8,s}\simeq\C^{r,s+8}\simeq M_{16}(\C^{r,s}),
\label{Bott}
\end{eqnarray}
\begin{eqnarray}
\C^{n+1,0}_{even}\simeq\C^{0,n+1}_{even}\simeq\C^{n,0}
\end{eqnarray}

The isomorphism (\ref{Cisom}) is given by
\begin{eqnarray}
\C^{r+1,s+1}&\ra& M_{2}(\R)\otimes \C^{r,s}\\
e_1&\ra&\sigma_1\otimes e_1\\
e_2&\ra&\sigma_3\otimes e_1\\
e_{i+1}&\ra&1_2\otimes e_i~~~(i=2,\dots,r)\\
e_{r+2}&\ra&\epsilon\otimes e_1\\
e_{r+2+i}&\ra&1_2\otimes e_{r+i}~~~(i=1,\dots,s),
\end{eqnarray}
where $\epsilon=i\sigma_2$.
The isomorphism (\ref{Bott}) is given by
\begin{eqnarray}
\C^{r+8,s}&\ra& M_{16}(\R)\otimes \C^{r,s}\\
e_i&\ra&\gamma_9^i\otimes e_1~~~(i=1,\dots,9)\\
e_{i+8}&\ra&1_{16}\otimes e_i~~~(i=2,\dots,r+s)
\end{eqnarray}
where $\gamma_9^i$ ($i=1,\dots,9$) are $SO(9)$ gamma matrices
represented as real symmetric $16\times 16$ matrices.

{}From these relations, it is sufficient to know
$\C^{n,0}$ and $\C^{0,n}$ ($n=1,2,\dots,8$)
to obtain the others. They are listed in table \ref{Clifford}.
\begin{table}[htb]
\begin{center}
$$
\begin{array}{c|ccc|c}
\hline\hline
n& \C^{n,0}&\C^{0,n}&\C_{even}^{n,0}\simeq\C_{even}^{0,n}&
\mbox{Gauge group}\\
\hline
1&\C&\R\oplus\R&\R&O\\
2&\H&M_2(\R)&\C&U\\
3&\H\oplus\H&M_2(\C)&\H&Sp\\
4&M_2(\H)&M_2(\H)&\H\oplus\H&Sp\times Sp\\
5&M_4(\C)&M_2(\H)\oplus M_2(\H)&M_2(\H)&Sp\\
6&M_8(\R)&M_4(\H)&M_4(\C)&U\\
7&M_8(\R)\oplus M_8(\R)&M_8(\C)&M_8(\R)&O\\
8&M_{16}(\R)&M_{16}(\R)&M_8(\R)\oplus M_8(\R)&O\times O\\
\hline
\end{array}
$$
\parbox{75ex}{
\caption{
\small Clifford algebra
}
\label{Clifford}
}
\end{center}
\end{table}

As an example,
let us consider $\C^{0,3}$, which is
used in section \ref{C03}. The algebra $\C^{0,3}$
is generated by $e_1$, $e_2$ and $e_3$ satisfying
(\ref{rel1})$\sim$(\ref{rel3}).
They are faithfully represented by Pauli matrices as
$e_1=\sigma_1$, $e_2=\sigma_2$ and $e_3=\sigma_3$.
Each element $A\in\C^{0,3}$ is of the form
\begin{eqnarray}
A&=&a+b e_1+ c e_2+ d e_3 + e e_1e_2 + f e_1e_3 + g e_2e_3 + h e_1e_2e_3\\
&=&\mat{a+hi+d+ei,b+gi-(f+ci),b+gi+f+ci,a+hi-(d+ei)},
\label{M2}
\end{eqnarray}
where $a,b,\dots,f\in\R$.
Since (\ref{M2}) is a general element of the algebra of
complex $2\times 2$ matrices $M_2(\C)$, we obtain
$\C^{0,3}\simeq M_2(\C)$.
The even elements are
\begin{eqnarray}
A_{even}&=&a+ e e_1e_2 + f e_1e_3 + g e_2e_3\\
&=&\mat{a+ei,-f+gi,f+gi,a-ei}
=a+gi\sigma_1-fi\sigma_2+ei\sigma_3.
\end{eqnarray}
Since $i\sigma_i$ ($i=1,2,3$)
can be represented by the generators of quaternion as
$i\sigma_1\ra i$, $i\sigma_2\ra j$ and $i\sigma_3\ra -k$,
we obtain $\C^{0,3}_{even}\simeq \H$.

Note that
we can immediately read the gauge groups of the type I
unstable D-brane systems from table \ref{Clifford}.
As explained in section \ref{realKKvsD}, the gauge groups
consist of unitary operators which are
 real operators tensored by even elements
of the corresponding Clifford algebra.
Recall that a unitary matrix whose matrix elements
are $\R$, $\C$ or $\H$-valued is an
orthogonal, unitary or symplectic matrix, respectively.
(See Appendix \ref{App2} for the symplectic case.)
Therefore, the gauge group of the system is $O$, $U$ or $Sp$,
 when the even elements of
the corresponding Clifford algebra is $M_n(\R)$, $M_n(\C)$ or
$M_n(\H)$, respectively,
as listed in the last column of table \ref{Clifford}.

\section{Quaternionic representation of $Sp$ group}
\label{App2}

Let us recall the definition of the $Sp$ group.\footnote{
The $Sp$ group we consider in this paper is
unitary symplectic group, which is often denoted
by $USp(2N)$.}
The (unitary) $Sp$ group consists of matrices
$g\in M_{2N}(\C)$ satisfying
\begin{eqnarray}
g^\dag g=1,~~~g^T\J g=\J
\label{USp}
\end{eqnarray}
where
\begin{eqnarray}
\J=\mat{,1,-1,}
\end{eqnarray}
(\ref{USp}) is equivalent to the condition that
$g$ is of the form
\begin{eqnarray}
g&=&g_a+g_b i\sigma_1+ g_c i\sigma_2+ g_d i\sigma_3
\label{g}\\
&=&\mat{g_a+ig_d,g_c+ig_b,-g_c+ig_b,g_a-ig_d },
\end{eqnarray}
satisfying $g^\dag g=1$, where $g_a$, $g_b$, $g_c$ and $g_d$
are real $N\times N$ matrices.

It is useful to represent $i\sigma_i$,
by the generators of quaternion as
$i\sigma_1\ra i$, $i\sigma_2\ra j$ and $i\sigma_3\ra -k$.
Then (\ref{g}) can be represented as
a quaternionic matrix
\begin{eqnarray}
g=g_a+g_b i+g_c j- g_d k\in M_N(\H).
\end{eqnarray}
In this way, the $Sp$ group can be represented as unitary
quaternionic matrices.

The adjoint representation
of the $Sp$-group is equivalent to the symmetric tensor
representation $\sym$. It is given by the anti-Hermitian matrices
$X\in M_{2N}(\C)$ satisfying
\begin{eqnarray}
(\J X)^T=\J X.
\label{sym}
\end{eqnarray}
The action of a element of 
the group $g$ is given by $X \rightarrow 
g^\dagger X g$.
In the quaternionic representation,
this condition is equivalent as saying that
$X\in M_N(\H)$ is quaternionic anti-Hermitian matrix.
Though this statement can be shown explicitly using (\ref{sym}),
it is obvious from the consideration above
since the adjoint representation is
given by the Lie algebra of the group.

The anti-symmetric tensor representation $\asym$ is
given by Hermitian matrices $X\in M_{2N}(\C)$ satisfying
\begin{eqnarray}
(\J X)^T=-\J X,
\label{asym}
\end{eqnarray}
which can be written as
\begin{eqnarray}
X=\mat{a+di,-b+ci,b+ci,a-di}= a+ bi\sigma_1+ ci\sigma_2+ di\sigma_3,
\label{X}
\end{eqnarray}
where $a,b,c,d\in M_N(\R)$ satisfy
$a=a^T$, $b=-b^T$, $c=-c^T$ and $d=-d^T$.
(\ref{X}) is represented as
\begin{eqnarray}
X= a+bi+cj-dk,~~~X^\dag=X
\end{eqnarray}
in the quaternionic representation.


\begin{thebibliography}{999}
\bibitem{MiMo}
R. Minasian and G. Moore,
``K-theory and Ramond-Ramond charge,''
JHEP {\bf 9711} (1997) 002,
hep-th/9710230.
\bibitem{Wi}E. Witten, ``D-branes and K-theory,'' JHEP {\bf 9812} (1998) 
 019, hep-th/9810188.
\bibitem{Ho}
P. Horava, ``Type IIA D-Branes, K-Theory, and Matrix Theory,''
JHEP {\bf 9901} (1999) 016, hep-th/9811028.
\bibitem{Se}
A. Sen,
``SO(32) Spinors of Type I and Other Solitons
on Brane-Antibrane Pair,''
JHEP {\bf 9809} (1998) 023, hep-th/9808141;
``Stable Non-BPS States and Branes in String Theory,''
hep-th/9904207, and references therein.
\bibitem{To}
P. K. Townsend,
``D-branes from M-branes,''
Phys. Lett. {\bf B373} (1996) 68,
hep-th/9512062.
\bibitem{BFSS}
T. Banks, W. Fischler, S. H. Shenker and L. Susskind,
``M Theory As A Matrix Model: A Conjecture,''
Phys. Rev. {\bf D55} (1997) 5112,
hep-th/9610043.
\bibitem{IKKT}
N. Ishibashi, H. Kawai, Y. Kitazawa and A. Tsuchiya,
``A Large-N Reduced Model as Superstring,''
Nucl. Phys. {\bf B498} (1997) 467,
hep-th/9612115.
\bibitem{Is}
N. Ishibashi,
``$p$-branes from $(p-2)$-branes
in the Bosonic String Theory,''
Nucl. Phys. {\bf B539} (1999) 107,
hep-th/9804163.
\bibitem{Kl}
J. Kluson,
``D-Branes from $N$ Non-BPS D0-Branes,''
JHEP {\bf 0011} (2000) 016, hep-th/0009189;
``D-Branes from $N$ D$(-1)$-Branes in Bosonic
and Type IIA String Theory,''
JHEP {\bf 0103} (2001) 018, hep-th/0102063.
\bibitem{Te}
S. Terashima, ``A Construction of Commutative D-branes from Lower Dimensional 
Non-BPS D-branes,''
JHEP {\bf 0105} (2001) 059,
hep-th/0101087 
\bibitem{AsSuTe}
T. Asakawa, S. Sugimoto and S. Terashima,
``D-branes, Matrix Theory and K-homology,''
hep-th/0108085.
\bibitem{Gu}
S. Gukov,
``K-Theory, Reality, and Orientifolds,''
Commun. Math. Phys. {\bf 210} (2000) 621, hep-th/9901042.
\bibitem{Kar}
M. Karoubi,
``K-Theory, An Introduction,''
Spriger-Verlag.
\bibitem{Be}
O. Bergman,
``Tachyon Condensation in Unstable Type I D-brane Systems,''
JHEP {\bf 0011} (2000) 015,
hep-th/0009252.
\bibitem{LoUr}
O. Loaiza-Brito and A. M. Uranga,
``The fate of the type I non-BPS D7-brane,''
Nucl. Phys. {\bf B619} (2001) 211,
hep-th/0104173.
\bibitem{HaMo}
J.A. Harvey and G. Moore, ``Noncommutative Tachyons and K-Theory,''
J. Math. Phys. {\bf 42} (2001) 2765,
hep-th/0009030.
\bibitem{Wi2}
E. Witten,
``Overview Of K-Theory Applied To Strings,''
Int. J. Mod. Phys. {\bf A16} (2001) 693,
hep-th/0007175.
\bibitem{KuMaMo}
D. Kutasov, M. Marino and G. Moore,
``Remarks on tachyon condensation in superstring field theory,'' 
hep-th/0010108.
\bibitem{KrLa}
P. Kraus and F. Larsen,
``Boundary String Field Theory of the DDbar System,''
Phys. Rev. {\bf D63} (2001) 106004, hep-th/0012198. 
\bibitem{TaTeUe}
T. Takayanagi, S. Terashima and T. Uesugi,
``Brane-Antibrane Action from Boundary String Field Theory,'' 
JHEP  {\bf 0003} (2001) 019, hep-th/0012210.
\bibitem{Kas}
G. G. Kasparov,
``The operator K-functor and extensions of $C^*$-algebras,''
Math. USSR Izv. {\bf 16} (1981) 513.
\bibitem{Bl}
B. Blackadar,
``K-Theory for Operator Algebras,''
Cambridge University Press.
\bibitem{Sc}
H. Schr\"oder,
``K-theory for real $C^*$-algebras and applications,''
Pitman Research Notes in Mathematics Series 290,
Longman Scientific and Technical.
\bibitem{Su}
S. Sugimoto,
``Anomaly Cancellations in the Type I
D$9$-D\protect$\ol 9$ System
 and the $USp(32)$ String Theory,''
Prog. Theo. Phys. {\bf 102} (1999) 685, hep-th/9905159.
\bibitem{Qu}
D. Quillen,
``Superconnection and the Chern Character,''
Topology {\bf 24} (1985) 89.
\end{thebibliography}
\end{document}